\documentclass[twocolumn]{IEEEtran}
\usepackage{cite}
\usepackage{amsmath,amssymb,amsfonts}
\usepackage{algorithmic}
\usepackage{graphicx}
\usepackage{textcomp}
\usepackage{soul}
\begin{document}
	
	\title{\bf A SVD-Based Synchrophasor Estimator for P-class PMUs with Improved Immune
		from Interharmonic Tones}
	
	\author{Dongfang Zhao, Fuping Wang, Shisong Li, \IEEEmembership{Senior Member, IEEE}, Lei Chen, \IEEEmembership{Member, IEEE}, \\ Wei Zhao, and Songling Huang, \IEEEmembership{Senior Member, IEEE} 
		\thanks{The authors are with the Department of Electrical Engineering, Tsinghua University, Beijing 100084, China.}
		\thanks{This work was supported in part by the National Natural Science Foundation of China under Grant No. 52077112, China Postdoctoral Science Foundation under Grant 2020M680551, and the Transforming Systems through Partnership under Grant No. TSPC1051 from the Royal Academy of Engineering, UK.}}

	\markboth{IEEE Access Manuscript}%
	{}
	
	\maketitle
	
	\begin{abstract}
		The increasing use of renewable generation, power electronic devices, and nonlinear loads in power systems brings more severe interharmonic tones to the measurand, which can increase estimation errors of P-class PMUs, cause misoperation of protection relays, and even threaten the stability of the power systems. Therefore, the performance of the P-class synchrophasor estimator under interharmonic interference should be evaluated and new estimation schemes that can improve the operational robustness are needed for various protection and control applications. In this paper, a synchrophasor estimator design for P-class PMUs that introduces singular value decomposition to the least square algorithm based on the Taylor series is proposed. By constructing an optimization with proposed adjustable parameters, finite impulse response filters with high transition band attenuation are designed. Numerical and experimental tests verify the proposed dynamic synchrophasor estimator performance and show an effective improvement in rejecting interharmonic tones, especially when a short time window and light computational burden are required. 
	\end{abstract}

	\begin{IEEEkeywords}
	Dynamic synchrophasor estimator, FIR filter, interharmonic, phasor measurement unit, singular value decomposition.
	\end{IEEEkeywords}
	\IEEEpeerreviewmaketitle
	
	\section{Introduction}
	\label{sec:introduction}
	\IEEEPARstart{T}{he} standard IEC/IEEE 60255-118-1:2018 \cite{relays2018118}, referred as the Standard below, does not set any requirement to reject interharmonic tones for P-class PMU algorithms. 
	However, numerous researches indicated that the interharmonic tones from the increasing use of renewable generation, power electronics devices and nonlinear loads could distort power system signals, leading to estimation errors for PMUs \cite{dervivskadic2017iterative,silva2021impact}. 
	This is especially crucial for P-class PMUs, because it may destruct the reactive power compensation devices, lead the protection relays to misoperate, and introduce unknown threatening to the stability of the power grids \cite{durna2016design}. It is therefore important to design synchrophasor estimators (SEs) that can immunize the various protection schemes, fault diagnosis, and control applications from interharmonic interference \cite{li2020second}.

	There are some dynamic synchrophasor algorithms compliment to P-class PMU requirements \cite{relays2018118}. The synchrophasor estimation based on enhanced phase-locked loop \cite{karimi2010application} reduced errors in single-phase frequency deviation and three-phase unbalanced conditions. The interpolated dynamic DFT-based SE \cite{petri2014frequency} could achieve high accuracy under off-nominal frequency and dynamic conditions. The least-square algorithms based on different basis functions \cite{chen2018dynamic,chen2020harmonic} were able to design filters with required passband width. The finite impulse response (FIR) filters designed with O-Splines \cite{de2020dynamic} could converge towards the ideal gain-frequency response. The optimal linear filter \cite{wang2018optiamlI} constructed the relationship between filter performance and estimation error. The dynamic SE using multiple frequency Taylor model \cite{fu2019multiple} could reduce the estimation error under dynamic modulations. However, these P-class PMU algorithms do not have sufficient attenuation to the out-of-band interference (OBI) for using a short window.
	
	Conventionally, for algorithms using a single signal model (containing only the fundamental tone), enlarging the window length helps to achieve high suppression of OBI. The response latency, however, will also increase \cite{petri2014frequency}. For example, some algorithms such as a dynamic phasor estimator considering OBI \cite{fu2021dynamic}, a dynamic phasor model-based method using two digital filters \cite{bi2015dynamic}, and a symmetric Taylor-based weighted least square algorithm \cite{radulovic2019dynamic} achieved the required accuracy under OBI at the cost of much longer response latency than P-class methods. Moreover, some P+M algorithms use two different configurations to respectively satisfy the P- and M-class PMU requirements, such as the DFT-based adaptive cascaded filters \cite{roscoe2013exploring}, Taylor-Fourier transform algorithm with different window parameters \cite{castello2014fast}, the band-pass synchrophasor filter based on complex brick-wall \cite{razo2019new}, and synchrophasor filters based on discrete-time frequency-gain transducer \cite{duda2021p}. They can realize either fast response for dynamic signal or high accuracy for steady signal, however, these algorithms can not achieve both features at the same time, especially for signals with OBI. 
	
	Recently, another idea, by introducing the interharmonic tones to the signal model, has been proposed to mitigate the negative influence of OBI under short window length. The iterative interpolated DFT (i-IpDFT) algorithm \cite{dervivskadic2017iterative} repeatedly estimated and compensated the spectral leakage of image fundamental and interharmonic tones, showing a good accuracy when dealing with signals contaminated by OBI. Its accuracy and computational burden are, however, inversely related to the iterative number. The matrix decomposition-based eigensystem realization \cite{zamora2020model} and Matrix Pencil \cite{li2020second,song2021accurate} set filters with notches at the interfering frequencies. Thus, they give a fast response and high accuracy when interharmonic interference is around the notch. The computational burden for synchrophasor estimation, however, can increase severely when there are many interharmonic tones.
	
	To this end, this paper investigates a different approach to improve the suppression of OBI, and at the same time to keep a low computational burden. The idea is to introduce the singular value decomposition (SVD) to the Taylor least square (TLS) algorithm, and realize an adjustable FIR filter and hence the optimized attenuation in the transition band, i.e., the OBI frequency range \cite{relays2018118}. The proposed SE can well balance the response latency and the measurement accuracy. Meanwhile, it does not need an iterative operation or a real-time matrix decomposition, and therefore will not cause heavy computation issues. For short, the proposed synchrophasor estimator based on SVD is denoted as SVDSE in the following text.
	
	The subsequent content of this paper is organized as follows: In section \ref{sec2}, the principle of the proposed SVDSE algorithm is presented. How to design the dynamic synchrophasor filter and optimize its attenuation in the transition band are shown in section \ref{sec3}. The gain-frequency response of the proposed SVDSE is compared to that of the TLS algorithm weighted by three different window functions, rectangular, Hanning, and Blackman, respectively. In section \ref{sec4}, numerical and experimental tests are carried out and the performance of the proposed SVDSE is discussed by comparing to TLS weighted by Blackman window and i-IpDFT algorithms. The frequency accuracy and computational burden analysis of SVDSE are shown in section \ref{sec5}. The conclusions are drawn in section \ref{sec6}.
	
	\section{Taylor Least Square Algorithm Based on Singular Value Decomposition}
	\label{sec2}
	\subsection{A brief recall of the TLS algorithm}
	\label{sec2A}
	
	In \cite{de2007dynamic}, the $K$-order Taylor expansion of dynamic synchrophasor $p(t)=a(t)\text{e}^{\text{j}\phi(t)}$ and its conjugation are used to represent the dynamic power system fundamental signal 
	\begin{align}
		s(t)=a(t)\cos(2\pi f_\text{r}t+\phi(t)), 
		\label{eq:1}
	\end{align}
	where $a(t)$, $f_\text{r}$, and $\phi(t)$ respectively denote the amplitude, reference frequency, and phase. The signal $s(t)$ is sampled with a frequency $f_\text{s}=1/T_\text{s}$ ($T_\text{s}$ is the sampling interval) and the sampling points in a time window $T_\text{w}$ are used as the input of the TLS algorithm. In order to avoid phase shift, the center of $T_\text{w}$ is regarded as initial time $t=0$. Then the time of each sample in the time window $T_\text{w}$ is $-N_hT_\text{s},\cdots,0,\cdots,N_hT_\text{s}$. Within $T_\text{w}$, the $N=2N_h+1$ samples yield the equations written in matrix \cite{de2007dynamic}, i.e.,
	\begin{align}
		\textbf{S}=\textbf{G}\textbf{P}=
		\begin{pmatrix}
			\textbf{E} &\textbf{E}^*
		\end{pmatrix}
		\begin{pmatrix}
			\textbf{B}_K &   \textbf{0}\\
			\textbf{0}   &   \textbf{B}_K
		\end{pmatrix}
		\textbf{P},
		\label{eq:2}
	\end{align}
	where $\textbf{S}\in\mathbb{R}^{N\times1}$ is a column vector with $N$ samples of the signal $s(t)$; $\textbf{G}\in\mathbb{C}^{N\times2(K+1)}$ is a matrix whose column contains $N$ samples of $\frac{t^k}{k!}\text{e}^{\text{j}2\pi f_\text{r}t}$ and its conjugation, i.e., its element $g_{ij}=\frac{((i-1-N_h)/f_\text{s})^{(j-1)}}{(j-1)!}\text{e}^{\text{j}2\pi f_\text{r}(i-1-N_h)/f_\text{s}}$  when $1\le i\le N$, $1\le j\le(K+1)$ and $g_{ij}=\frac{((i-1-N_h)/f_\text{s})^{(j-K-2)}}{(j-K-2)!}\text{e}^{-\text{j}2\pi f_\text{r}(i-1-N_h)/f_\text{s}}$ when $1\le i\le N, (K+2)\le j\le2(K+1)$; $\textbf{P}\in\mathbb{C}^{2(K+1)\times1}$ is a column vector formed by linear parameters to be solved, i.e., $p_0, p_1, ..., p_K$ in the $K$-order Taylor expansion of the dynamic synchrophasor $p(t)$ and their conjugations; $\textbf{E}\in\mathbb{C}^{N\times N}$ is a diagonal matrix and the diagonal elements are $N$ samples of $\text{e}^{\text{j}2\pi f_\text{r}t}$; $\textbf{B}_K\in\mathbb{R}^{N\times(K+1)}$ is a matrix using the $N$ samples of basic function $\frac{t^k}{k!}$ as the elements in $k$-th column; $^*$ represents the conjugate operation of matrix. 
	
	In (\ref{eq:2}), there are in total $L\times c$ equations, where $L$ and $c$ are respectively the number of samples in one nominal fundamental cycle $T_0$, i.e., $L=f_\text{s}T_0$, and the number of nominal fundamental cycles in one time window $T_\text{w}$, i.e., $c=T_\text{w}/T_0$. Suppose all signal components are expanded to the same order $K$. Therefore there are at most $2H(K+1)$ variables to be solved, where $2H$ represents the number of synchrophasor and its conjugate components. The maximum frequency of the signal components should be below half of the signal sampling frequency to avoid spectrum aliasing.  We note $(K+1)\neq c$ is often related to high response latency or computation instability for the proposed algorithm in section \ref{sec3}. Here we choose an easy and stable case, i.e. $(K+1)=c$. Moreover, (\ref{eq:2}) is usually an overdetermined system of linear equations, and its optimal solution can be solved by the least square method, i.e.,
	\begin{equation}
		\hat{\textbf{P}}=(\textbf{G}^{\text{H}}\textbf{G})^{-1}\textbf{G}^{\text{H}}\textbf{S}=\textbf{G}^+\textbf{S},
		\label{eq:3}
	\end{equation}
	where $^{\text{H}}$ denotes the conjugate transpose operation of matrix; $^{-1}$ and $^+$ respectively show the inverse and generalized inverse operations of matrix. The time-inverse elements of each row in the matrix $\textbf{G}^+$ are the FIR coefficients for solving the variables in the vector $\textbf{P}$.
	
	Using the dynamic synchrophasor $p(t)$ and its first and second derivatives, the formula for the amplitude, phase, frequency and rate of change of frequency (RoCoF) of the dynamic synchrophasor can be obtained as \cite{de2007dynamic}
	\begin{align}
		a(t)&=2|p_0|\nonumber\\
		\phi(t)&=\angle p_0\nonumber\\
		f(t)&=f_\text{r}+\frac{1}{2\pi}\frac{\text{Im}(p_1p_0^*)}{|p_0|^2}\nonumber\\
		f'(t)&=\frac{1}{2\pi}\left(\frac{\text{Im}(p_2p_0^*)}{|p_0|^2}-\frac{\text{2Re}(p_1p_0^*)\text{Im}(p_1p_0^*)}{|p_0|^4}\right),
		\label{eq:4}
	\end{align}
	where Re($\cdot$) and Im($\cdot$) respectively represent the real and imaginary part of a complex number. 
	
	\subsection{Introducing the SVD into the TLS algorithm}
	The singular value decomposition is often used to decompose the Hankel matrix formed by signal samples. It helps to reduce the negative influence of noise by preserving the large singular values \cite{kalman2018svd}. This effect enlightens us to enhance the filter's suppression of interharmonic tones and noise interference by introducing the SVD to the filter related matrix $\textbf{B}_K$ in (\ref{eq:2}) without increasing the time window length. The SVD of $\textbf{B}_K$ is given as follows \cite{kalman2018svd}  
	\begin{equation}
		\textbf{B}_K=\textbf{U}\Sigma\textbf{V}^\text{T},
		\label{eq:5}
	\end{equation}
	where matrix $\textbf{U}\in\mathbb{R}^{N\times (K+1)}$, and its column vector $\textbf{u}_i(1\le i\le K+1)$ is the so-called left singular vector; orthogonal matrix $\textbf{V}\in\mathbb{R}^{(K+1)\times (K+1)}$, and its column vector $\textbf{v}_i(1\le i\le K+1)$ is the right singular vector; $\Sigma \in\mathbb{R}^{(K+1)\times (K+1)}$ is a diagonal matrix, and its $(K+1)$ diagonal elements are the singular values. Substituting (\ref{eq:5}) into (\ref{eq:2}) and comparing it with (\ref{eq:3}), the matrix $\textbf{G}^+$ is written as
	\begin{align}
		\textbf{G}^+&=
		\begin{pmatrix}
			\textbf{V} &   \textbf{0}\\
			\textbf{0}   &   \textbf{V}
		\end{pmatrix}
		\begin{pmatrix}
			\Sigma^{-1} &   \textbf{0}\\
			\textbf{0}   &   \Sigma^{-1}
		\end{pmatrix}
		\nonumber\\&
		\begin{pmatrix}
			\textbf{I}_{K+1} &   \textbf{U}^{\text{T}}\textbf{E}^{2*}\textbf{U}\\
			\textbf{U}^{\text{T}}\textbf{E}^{2}\textbf{U}   &   \textbf{I}_{K+1}
		\end{pmatrix}
		^{-1}
		\begin{pmatrix}
			\textbf{U}^{\text{T}}\textbf{E}^{*}\\
			\textbf{U}^{\text{T}}\textbf{E}
		\end{pmatrix},
		\label{eq:6}
	\end{align}
	where $\textbf{E}^{2*}=(\textbf{E}^2)^*$; $\textbf{I}_{K+1}\in\mathbb{R}^{(K+1)\times (K+1)}$ is an identity matrix. Note the elements of the inverse matrix in (\ref{eq:6}) are conjugate symmetric about the main- and anti-diagonal, then it can be denoted as        
	\begin{equation}
		\begin{pmatrix}
			\textbf{C} &   \textbf{D}\\
			\textbf{D}^* &   \textbf{C}^*
		\end{pmatrix}=\begin{pmatrix}
			\textbf{I}_{K+1} &   \textbf{U}^{\text{T}}\textbf{E}^{2*}\textbf{U}\\
			\textbf{U}^{\text{T}}\textbf{E}^{2}\textbf{U}   &   \textbf{I}_{K+1}
		\end{pmatrix}^{-1},
		\label{eq:7}
	\end{equation}
	where matrices $\textbf{C}$, $\textbf{D}\in\mathbb{C}^{(K+1)\times(K+1)}$ are composed of matrices $\textbf{U}$ and $\textbf{E}$, i.e.,
	\begin{align}
		\textbf{C}&=(\textbf{I}_{K+1}-\textbf{U}^{\text{T}}\textbf{E}^{2*}\textbf{U}\textbf{U}^{\text{T}}\textbf{E}^2\textbf{U})^{-1}\nonumber\\
		\textbf{D}&=-\textbf{C}\textbf{U}^{\text{T}}\textbf{E}^{2*}\textbf{U}.
		\label{eq:8}
	\end{align}
	Using (\ref{eq:6}) and (\ref{eq:7}), matrix $\textbf{G}^+$ can be rewritten as
	
	\begin{align}
		\textbf{G}^+&=
		\begin{pmatrix}
			\textbf{V}\Sigma^{-1}\textbf{C}\textbf{U}^{\text{T}}\textbf{E}^{*}\textbf{L}\\
			\textbf{V}\Sigma^{-1}\textbf{C}^*\textbf{U}^{\text{T}}\textbf{E}\textbf{L}^*
		\end{pmatrix}=
		\begin{pmatrix}
			\textbf{G}_\text{p}^+\\
			\textbf{G}_\text{n}^+
		\end{pmatrix},
		\label{eq:9}
	\end{align}
	where $\textbf{L}=(\textbf{I}_N-\textbf{E}^{*}\textbf{U}\textbf{U}^{\text{T}}\textbf{E})$, identity matrix $\textbf{I}_{N}\in\mathbb{R}^{N\times N}$, $\textbf{G}_\text{p}^+$, $\textbf{G}_\text{n}^+\in\mathbb{C}^{(K+1)\times N}$. The time-inverse elements in each row of matrix $\textbf{G}_\text{p}^+$ are the FIR coefficients of filters for dynamic synchrophasor and its derivatives, whereas that of matrix $\textbf{G}_\text{n}^+$ are the FIR coefficients for their conjugations. 
	
	\section{Design of Dynamic Synchrophasor Filter with improved transition band performance}
	\label{sec3}
	
	In this section, the design of proposed dynamic synchrophasor filter is presented. It can be seen from (\ref{eq:8}) and (\ref{eq:9}) that matrices $\textbf{C}$ and $\textbf{L}$ are determined by $\textbf{U}$ and $\textbf{E}$. Therefore, in (\ref{eq:9}), only matrices $\textbf{U}$, $\textbf{E}$, $\textbf{V}$, $\Sigma$ can be changed independently to alter the filter's gain-frequency response. In this respect, we investigate an easy possibility of adjusting the four matrices in (\ref{eq:9}) and $\textbf{V}\Sigma^{-1}$ is chosen as the adjustable part to improve the gain-frequency of dynamic synchrophasor filter. The reasons are: 1) The matrix $\textbf{E}$ is a frequency shifter and can not change the transition band performance; 2) $\textbf{U}$ is a large matrix and its adjustment is complicated because it affects other variables, such as $\textbf{C}$ and $\textbf{L}$; 3) the weight pair, $\textbf{V}\Sigma^{-1}$, is a simple factor to adjust and can turn freely the gain-frequency response. In (\ref{eq:9}), with $\textbf{V}\Sigma^{-1}$ adjustable, the time-inverse version for FIR coefficients of designed dynamic synchrophasor filter, $\textbf{h}_1\in\mathbb{C}^{1\times N}$ is written as
	\begin{align}
		\textbf{h}_1=\sum_{k=1}^{K+1}\frac{m_{1k}v_{1k}}{\sigma_{k}}\textbf{r}_k,
		\label{eq:10}
	\end{align}
	where $\textbf{r}_k\in\mathbb{C}^{1\times N}(1\le k\le K+1$) denotes the $k$-th row elements of matrix $\textbf{R}=\textbf{C}\textbf{U}^{\text{T}}\textbf{E}^{*}\textbf{L}$; $v_{1k}$ is the $k$-th element of the first row in right singular matrix $\textbf{V}$; $\sigma_{k}$ is the $k$-th element in the diagonal of singular value matrix $\Sigma$; $m_{1k}$ is the introduced multiplier to the weight $v_{1k}/\sigma_{k}$. 
	
	Eq. (\ref{eq:10}) shows that $\textbf{h}_1$ is the weighted sum of row elements in $\textbf{R}$. When all $m_{1k}$ values equal to one, (\ref{eq:10}) gives the original filter FIR coefficients to be optimized. It is shown in the appendix that the even elements in the first row of $\textbf{V}$, $v_{1(2a)}(1\le a\le\lfloor(K+1)/2\rfloor\text{,}\enskip\lfloor\cdot\rfloor \text{ represents the round down operation})$ equal zero, and the odd elements $v_{1(2a+1)}(0\le a\le\lfloor K/2\rfloor)$ are non-zero numbers. As a result, only the odd multipliers, $m_{1(2a+1)}$ attached to $v_{1(2a+1)}$, can be used to change the FIR coefficients of dynamic synchrophasor filter, and the even multipliers,  $m_{1(2a)}$ are set to be one. 
	
	Given the relationship between the effective multipliers $m_{1(2a+1)}$ and filter FIR coefficients or the gain-frequency response, an optimal problem is constructed to design the dynamic synchrophasor filter with high attenuation in transition band, i.e.,
	\begin{align}
		\mathop{\textbf{M}\text{i}\text{n}}\limits_{\substack{m_{1(2a+1)} \\0\le a\le\lfloor K/2\rfloor}}\{\mathop{\text{max}}\limits_{j\in \Omega}(H_{1j})\}\nonumber\\
		\textbf{s.}\textbf{t.} \quad \forall \quad m_{1(2a+1)}>0 \nonumber\\
		\exists\quad m_{1(2a+1)}\neq1,
		\label{eq:11}
	\end{align}
	where: $H_{1j}$ is the Fourier transform of the time-inverse of $\textbf{h}_1$ in (\ref{eq:10}), representing the filter gain at frequency point $j$; $\Omega$ denotes the frequency range for the OBI, i.e., the transition band. The optimal objective is to minimize the maximum filter gain in the transition band by changing the multipliers $m_{1(2a+1)}$. The first constrain is used to keep the computation stability, and the second one means that at least one multiplier is changed compared to the original ($m_{1(2a+1)}=1$). As defined in section \ref{sec2A}, the window length $c$ (number of fundamental periods) depends on the signal's Taylor expansion order $K$, i.e., $c=K+1$. When $c=1$ or $c=2$, only one effective multiplier $m_{11}$ is usable for adjusting the filter performance. In this case, it is hard to achieve a good dynamic synchrophasor estimator, especially when the fundamental component is contaminated by interharmonic tones. When $c$ is large, e.g. $c\ge4$, more adjustable multipliers are optional. However, the response latency will increase and the algorithm becomes unsuitable for real-time applications. Whereas $c=3$ yields two variables, $m_{11}$ and $m_{13}$, to optimize the dynamic performance of the synchrophasor filter. With such a configuration, the filter's response latency is short and a high performance in the transition band is achievable, therefore, in the proposed SVDSE, $c$ is set to 3.
	
\begin{figure}
		\centering
		\includegraphics[width=0.5\textwidth]{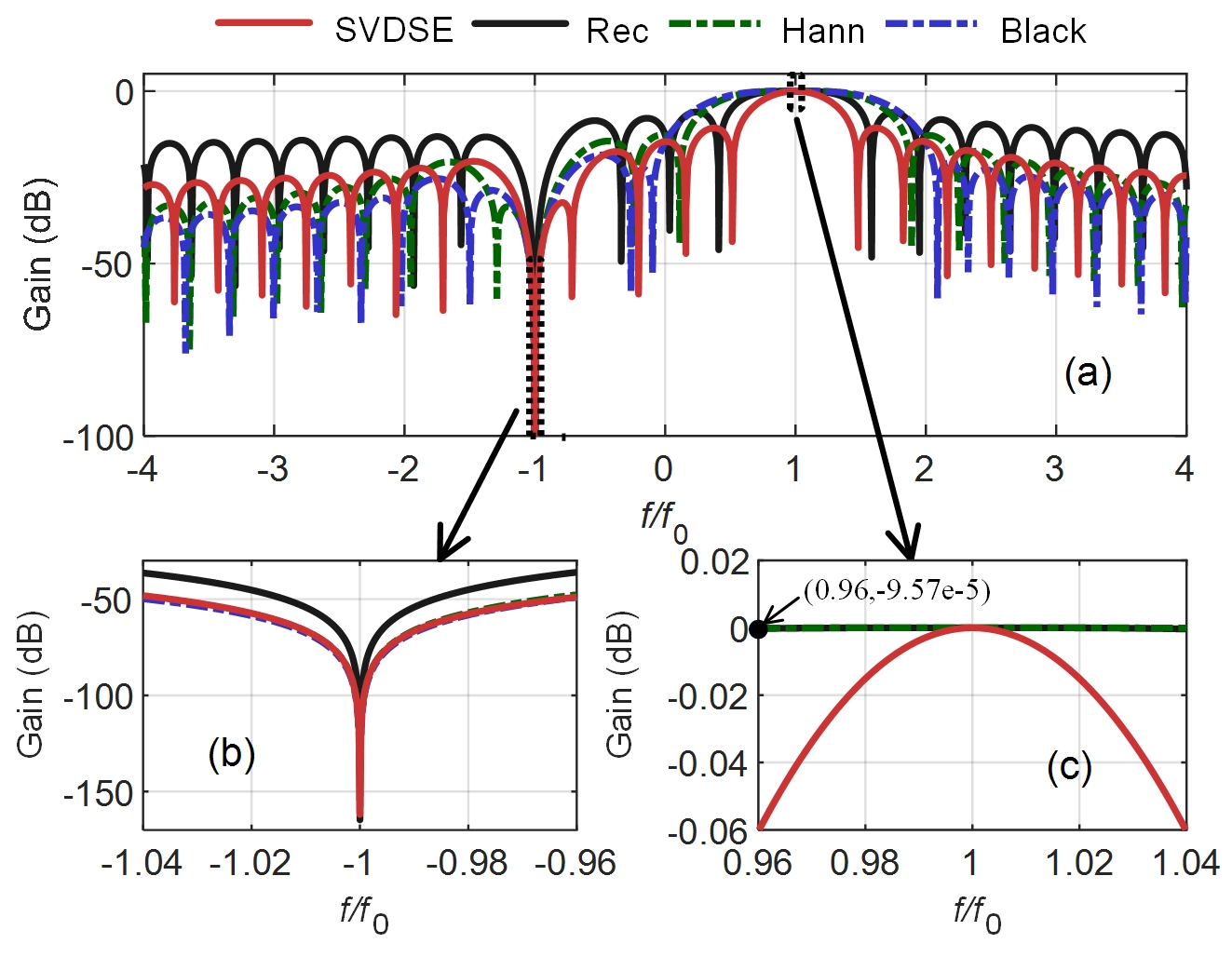}
		\caption{The gain-frequency responses of dynamic synchrophasor filters obtained by SVDSE and TLS weighted by rectangular (Rec), Hanning (Hann), and Blackman (Black) window functions [see (a)]. (b) and (c) are zooms around negative and positive nominal frequency, respectively.}
		\label{fig:01Filter}
	\end{figure}

	Numerous simulations show that when $m_{11}\neq1$, the passband ($\pm 2$ Hz for P-class PMU) ripple of the designed filter is unacceptable, and therefore $m_{11}$ is kept one. By contrast, increasing $m_{13}$ yields that the maximum gain in the transition band decreases first and then increases. As a result, it is obtained that when $m_{11}=1$ and $m_{13}=2.44$, the designed synchrophasor filter realizes the best transition band performance. As a comparison, the gain-frequency responses of filters obtained by SVDSE and TLS respectively weighted by rectangular, Hanning, and Blackman window functions are shown in Fig. \ref{fig:01Filter}. It can be seen that among the four filters, the dynamic synchrophasor filter designed by the proposed SVDSE has the highest attenuation in most of the transition band, $f/f_0\in(0,0.5)$ and $ (1.5,2)$, where $f_0$ is the nominal frequency. In contrast, the filtering gain in transition band of TLS algorithm weighted by other windows is greater than rectangular, resulting in worse suppression of interharmonic tones.
	
	As is seen from Fig. \ref{fig:01Filter}, the passband ripple of the proposed dynamic synchrophasor filter increases, lowering the sensitivity to track signals with frequency deviation or dynamic characteristics. It is shown that exact frequency reference helps to achieve accurate synchrophasor measurement \cite{zhan2016clarke}. Therefore, we set the reference frequency in $\textbf{E}$ to the estimated frequency in the previous time window to suppress the passband ripple effect.
	
	In summary, the off-line optimal and on-line application processes of the proposed SVDSE are shown in table \ref{tab:01}. Note that the proposed method can also be used to improve the transition band performance of the $i$-th synchrophasor derivative filter, e.g., the first and second derivatives used in solving frequency and RoCoF. However, the constructed multipliers for optimizing $i$-th synchrophasor derivative, $m_{(i+1)k}\,(1\le i\le K,\, 1\le k\le K+1)$ are different from that for synchrophasor, $m_{1k}\,(1\le k\le K+1)$ in (10). For example, the RoCoF accuracy under interharmonic tones depends not only on $m_{1k}$, but also on $m_{2k}$ and $m_{3k}$. In this paper, optimizing only $m_{1k}$ while keeping $m_{2k}$ and $m_{3k}$ unchanged does not affect much the RoCoF accuracy. For readability, the optimization of synchrophasor derivative filters' performance is not discussed in this paper.
	
	\begin{table}[tp!]
		\centering
		\caption{The optimal and application process of SVDSE.}
		\setlength{\tabcolsep}{1.8mm}
		\renewcommand\arraystretch{1.5}
		\begin{tabular}{p{220pt}}
			\hline
			\textbf{1) The off-line optimal design of SVDSE}\\
			\hline
			i) Obtain the matrix $\textbf{B}_K$ composed of the samples of basic functions shown in (\ref{eq:2})\\
			ii) Perform the SVD to matrix $\textbf{B}_K$ as (\ref{eq:5}) shows to obtain the weight product pair $\textbf{V}\Sigma^{-1}$\\
			iii) Distinguish the effective multipliers as the appendix shows, construct and solve the optimal problem shown in (\ref{eq:11}) by numerical research to obtain the best values of effective multipliers\\
			\hline
			\textbf{2) The on-line application part of SVDSE}\\
			\hline
			i) For the first data frame, the reference frequency in $\textbf{E}$ is set to nominal value. Then use the optimized multipliers to design filters shown in (\ref{eq:10}), compute the elements in \textbf{P} by (\ref{eq:3}), and estimate the synchrophasor and frequency using (\ref{eq:4})\\
			ii) For the following data frame, let the estimated frequency at last frame as the reference frequency in $\textbf{E}$ at current data frame. The following steps are the same as last phase\\
			\hline
		\end{tabular}
		\label{tab:01}
	\end{table}
	
	\section{Numerical and Experimental Tests}
	\label{sec4}
	
	\subsection*{1) Numerical tests}
	
	In this section, a set of typical steady, dynamic, and transient tests referring to the Standard \cite{relays2018118} are used to evaluate the performance of the proposed SVDSE algorithm. For comparison, other two commonly used or state-of-the-art algorithms, i.e., weighted TLS (TWLS) \cite{platas2010dynamic} and i-IpDFT \cite{dervivskadic2017iterative} are also tested. The TWLS algorithm uses the Blackman window function with good sidelobe attenuation \cite{hampannavar2020performance} and the i-IpDFT applies Hanning window function \cite{dervivskadic2017iterative} having maximum sidelobe decay within two-term cosine windows. 
	
	The definition of total vector error (TVE), response time, and reporting rate defined in \cite{relays2018118} is used in these tests. Some general test settings are: the sampling frequency $f_\text{s}=5000\,$Hz for SVDSE and TWLS whereas $f_\text{s}=50000\,$Hz for i-IpDFT \cite{dervivskadic2017iterative}; the time window $T_{\text{w}}=3/f_0$ and the reporting rate $f_{\text{re}}=50\,$frames/sec for all three algorithms (It is worth mentioning that in the Standard, four different frame rates, $f_{\text{re}}=\,$10, 25, 50, and 100 frames/sec, are used. For the proposed SVDSE algorithm, a minor change of the test results is obtained under these reporting rates.); the Taylor expansion order $K=2$ for SVDSE and TWLS; the multipliers $m_{11}=1$, $m_{12}=1$, and $m_{13}=2.2$ for SVDSE (smaller $m_{13}$ than optimization 2.44 helps to satisfy the response time limits required by the Standard); the iterative numbers for mitigating the negative influence of image tones and interharmonics are 2 and 28 for i-IpDFT, respectively, and the threshold for judging whether there is interharmonic is set to $\alpha=3.3\times10^{-3}$ \cite{dervivskadic2017iterative}. The test conditions and index limits of TVE for P-class PMUs and OBI test of M-class PMUs in \cite{relays2018118} are referred in the following tests.
	
	The filter performance of synchrophasor and its derivatives is improved by different multipliers as discussed in the last paragraph of section \ref{sec3}. Since the focus of this paper is optimizing the filter's transition band performance for synchrophasor instead of its derivatives, the error of unoptimized frequency and RoCoF, i.e. FE and RFE defined in \cite{relays2018118} will not be used as the evaluation indices in the following tests. In section \ref{sec5}, the FE, however, is applied to verify that the frequency adaption method in SVDSE is effective and the obtained frequency does not diverge.
	
	\textit{Case A: OBI with different amplitude.} The input signal for OBI test is given as
	\begin{equation}
		x(t)=\cos (2\pi ft+\phi_1)+A_\text{i}\cos (2\pi f_\text{i}t+\phi_\text{i}).
		\label{eq:12}
	\end{equation}
	For every test, only one interharmonic tone is added. The fundamental frequency $f=50\,$Hz and interharmonic frequency is set to its nearest $25\,$Hz among the OBI frequency interval [10, 25]$\,$Hz in the Standard when reporting rate is $50\,$ frames/sec (similar result appears in the other interval [75, 100]$\,$Hz). The phases of the fundamental $\phi_1$ and the interharmonic $\phi_\text{i}$ are randomly chosen within $[-\pi, \pi]$. The amplitude of the interharmonic tone $A_\text{i}$ ramps from 1\% to 20\% of the fundamental with a step of 1\%. 
	
\begin{figure}
		\centering
		\includegraphics[width=0.5\textwidth]{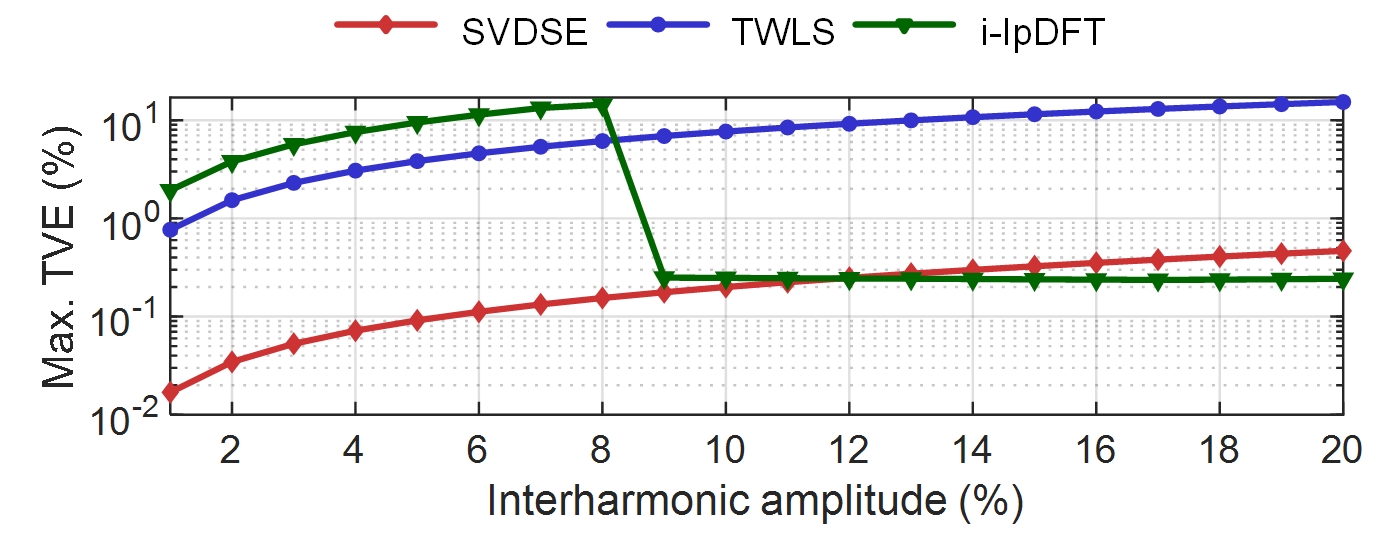}
		\caption{Maximum TVE versus interharmonic amplitude obtained by SVDSE, TWLS, and i-IpDFT algorithms under OBI condition.}
		\label{fig:02intera}
	\end{figure}

	The maximum TVE varying with the interharmonic amplitude is shown in Fig. \ref{fig:02intera}. When the interharmonic amplitude is small, e.g., under 11\% of the fundamental tone, the proposed SVDSE algorithm has the lowest TVE. A turning point is observed in the i-IpDFT performance with the increasing interharmonic amplitude. The reason is that the detection sensitivity of the i-IpDFT algorithm is limited and can not detect interharmonics with low amplitudes. Although the detection sensitivity can be increased by lowering the judgment threshold, however, the risk is increasing the possibility to mix between a dynamic behavior (effective signal) and an interharmonic tone to be eliminated \cite{dervivskadic2017iterative}. By contrast, the TVE obtained by TWLS is high in the whole range, because other window functions have greater transition band gain than the rectangular window as shown in Fig. \ref{fig:01Filter}, thus weakening the suppression of interharmonic interference.
	
	\textit{Case B: OBI with different frequency.} The signal model in (\ref{eq:12}) is also used in this test. Similarly, only one interharmonic tone is added in every test. There is a minus fundamental frequency deviation $-2\,$Hz from the nominal value, i.e., $f=48\,$Hz because a frequency lower than nominal might suffer even more from spectral leakage effects when using a short window length. The interharmonic amplitude $A_\text{i}$ is 10\% of the fundamental and its frequency ramps within two frequency intervals [10, 25]$\,$Hz and [75, 100]$\,$Hz with a step of $2.5\,$Hz. The fundamental and the interharmonic phases are random numbers within $[-\pi, \pi]$.
	
\begin{figure}
		\centering
		\includegraphics[width=0.5\textwidth]{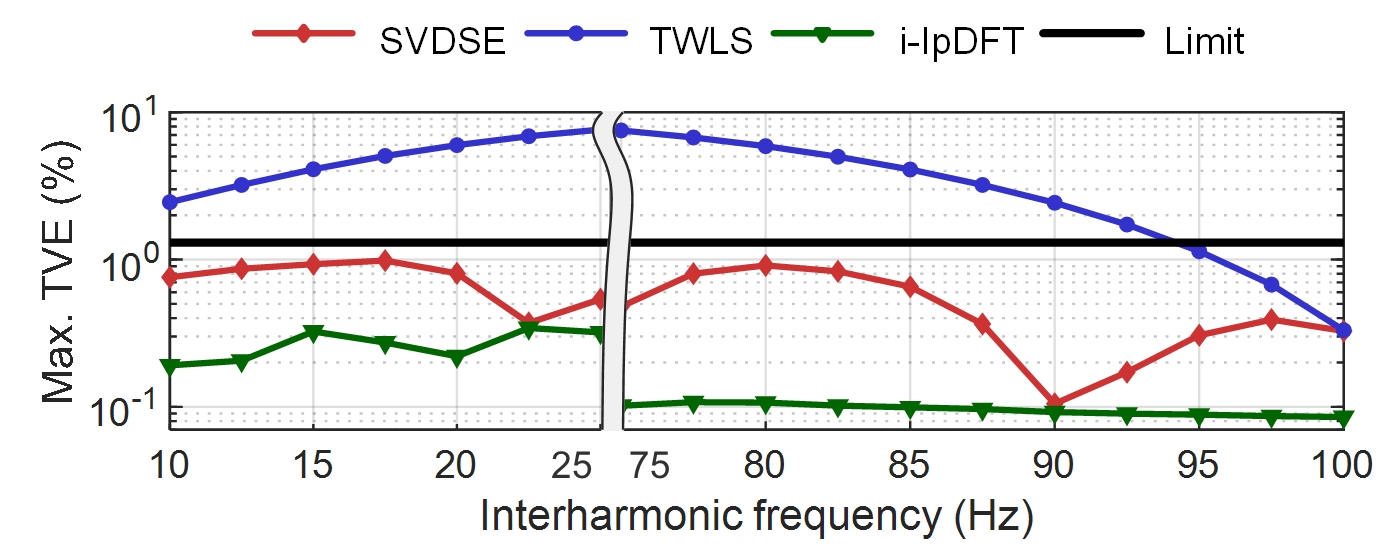}
		\caption{Maximum TVE versus interharmonic frequency obtained by SVDSE, TWLS, and i-IpDFT algorithms under OBI condition.}
		\label{fig:03interf}
	\end{figure}

	Fig. \ref{fig:03interf} illustrates the maximum values of TVE obtained by the SVDSE, TWLS, and i-IpDFT algorithms and the limit in the Standard (1.3\%). It can be seen that the TVE of the proposed SVDSE is within the limit in the test range, and is slightly higher than that obtained by the i-IpDFT. Whereas, the TVE of TWLS exceeds the limit in most ranges. The test results agree with the prediction by observing the transition band performance shown in Fig. \ref{fig:01Filter} and verify that the proposed SVDSE is less sensitive to interharmonics.
	
	\textit{Case C: Noise interference and OBI.} In this test, a white noise $u(t)$ and an interharmonic tone are added to the fundamental signal, i.e.,
	\begin{equation}
		x(t)=\cos (2\pi ft+\phi_1)+u(t)+0.05\cos (2\pi f_\text{i}t+\phi_\text{i}).
		\label{eq:13}
	\end{equation}
	The fundamental has the frequency $f=52\,$Hz and a random phase within $[-\pi, \pi]$. The signal-to-noise (SNR) of $x(t)$ is set from 40 to 80\,dB \cite{bhandari2019real} with a step of 5\,dB. The interharmonic frequency is set to $20\,$Hz that is beneficial to i-IpDFT and harmful to proposed SVDSE, as shown in the last test. The interharmonic amplitude is set to 5\% of the fundamental and its phase is randomly chosen within $[-\pi, \pi]$. Note that in the following tests from case D to G, the contained interharmonic tone has the same settings as this test, and it will not be repeatedly stated.
	
\begin{figure}
		\centering
		\includegraphics[width=0.5\textwidth]{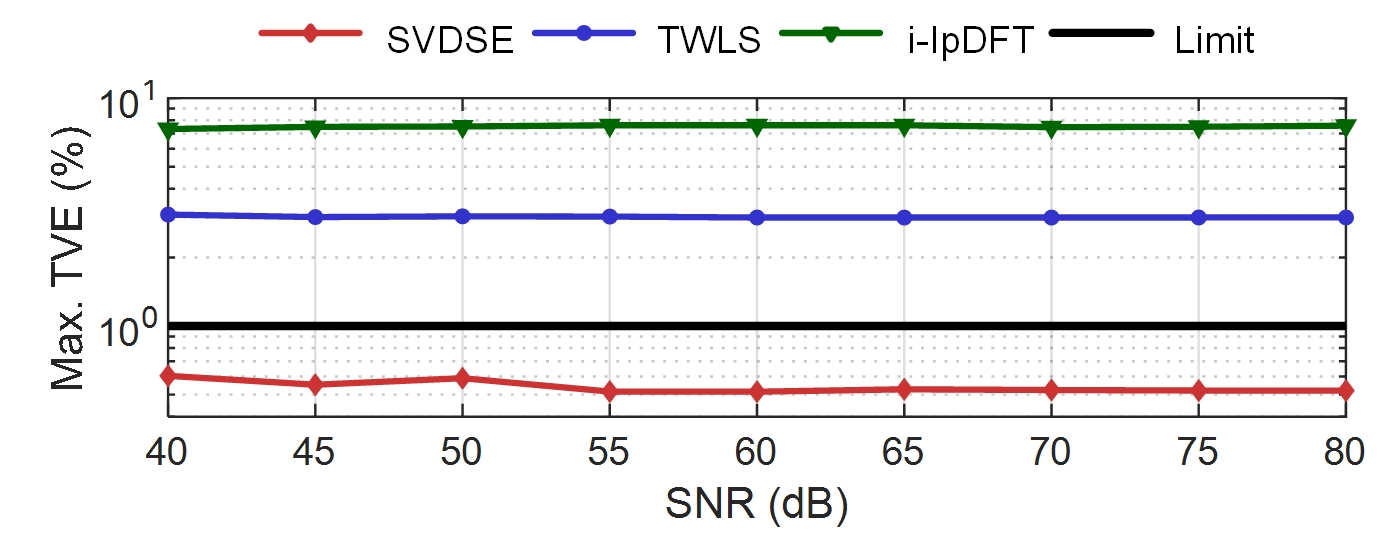}
		\caption{Maximum TVE versus SNR obtained by SVDSE, TWLS, and i-IpDFT algorithms under noise interference and OBI condition.}
		\label{fig:04noise}
	\end{figure}

	The TVE under different SNR setups is analyzed using the three algorithms. The maximum values and the TVE limit which is supposed to be 1\% are shown in Fig. \ref{fig:04noise}. It is observed that the proposed SVDSE algorithm exhibits the best noise immunity over the three algorithms and the TVE is still well below the limit with the SNR decreasing to $40\,$dB. The result relies on the fact that the optimal multipliers of the proposed filter help to increase the attenuation out of the passband.   
	
	\textit{Case D: Harmonic interference and OBI.} The test signal contains the fundamental, $h$-th$~(2\le h\le 50)$ order harmonic with the amplitude of 1\% of the fundamental \cite{relays2018118}, and one interharmonic tone, i.e.,
	\begin{equation}
		\begin{split}
			x(t)=\cos (2\pi ft+\phi_1)+0.01\cos (2\pi hft+\phi_\text{h})\\
			+0.05\cos (2\pi f_\text{i}t+\phi_\text{i}).
		\end{split}
	\end{equation}
	The fundamental frequency $f=50$\,Hz, and its phase $\phi_1$ and harmonic phase $\phi_\text{h}$ are random numbers within $[-\pi,\pi]$.
	
\begin{figure}
		\centering
		\includegraphics[width=0.5\textwidth]{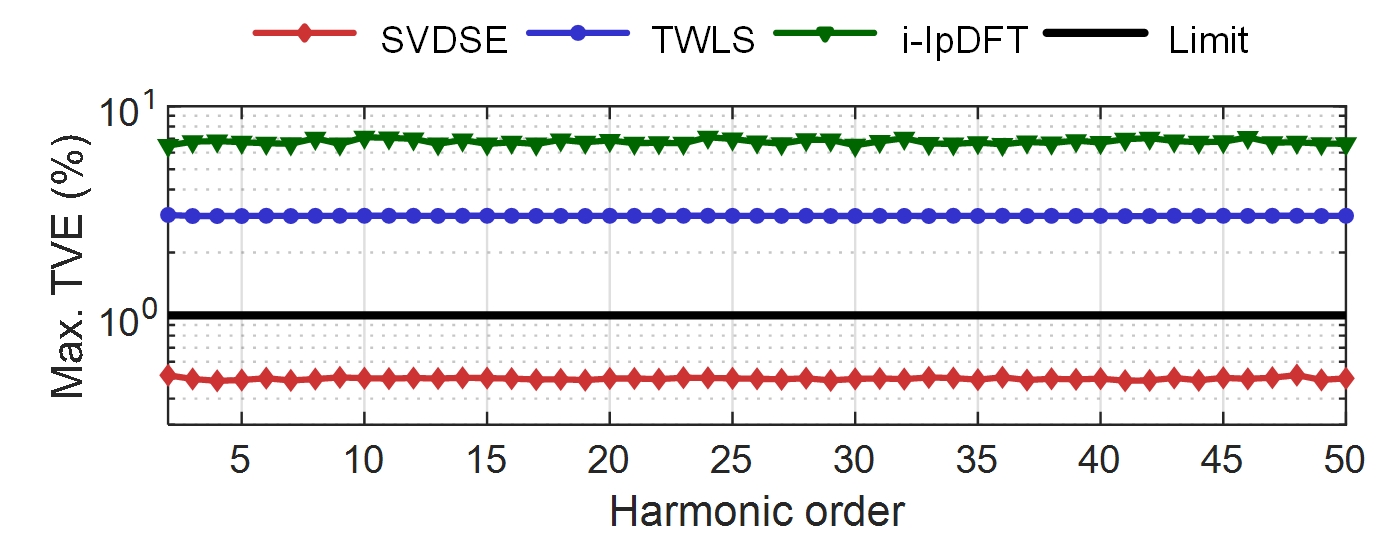}
		\caption{Maximum TVE versus harmonic order obtained by SVDSE, TWLS, and i-IpDFT algorithms under harmonic interference and OBI condition.}
		\label{fig:05har}
	\end{figure}

	The maximum TVE versus harmonic order obtained by the three algorithms and the limit, i.e., 1\%, are shown in Fig. \ref{fig:05har}. Note the TVE of the proposed SVDSE is also well below the limit and has a better suppression of harmonics and interharmonics than i-IpDFT and TWLS. 
	
	\textit{Case E: Frequency deviation and OBI.} The signal used in this test contains the fundamental and interharmonic components with a form of (\ref{eq:12}). The fundamental frequency $f$ increases from 48\,Hz to 52\,Hz with a step of 0.1\,Hz and its phase $\phi_1$ is a random number within $[-\pi, \pi]$. 
	
\begin{figure}
		\centering
		\includegraphics[width=0.5\textwidth]{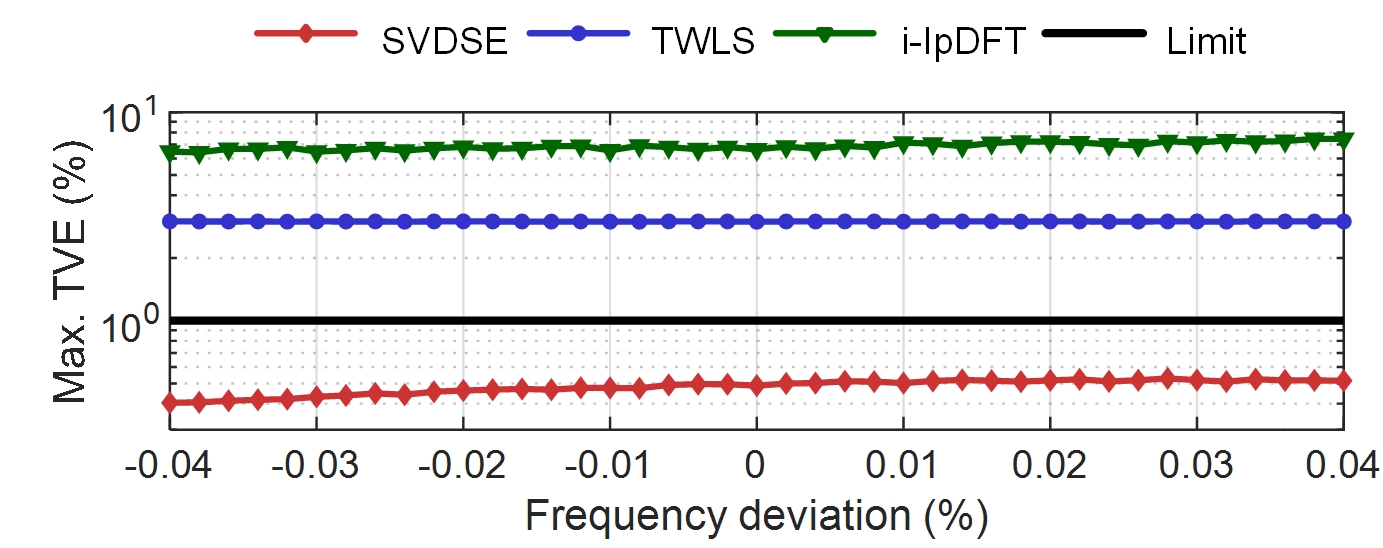}
		\caption{Maximum TVE versus frequency deviation obtained by SVDSE, TWLS, and i-IpDFT algorithms under frequency deviation and OBI condition.}
		\label{fig:06fdev}
	\end{figure}

	As shown in Fig. \ref{fig:06fdev}, the maximum TVE of SVDSE is the lowest among the three algorithms and well within the limit (1\%) for this test. The results show that using the dynamic signal model, and setting the center frequency for every frame to the estimated frequency of the previous data window help to mitigate the negative effects of frequency deviation. Note that the frequency adaption does not increase much computation and the execution time is short enough to meet the real-time reporting, which shall be analyzed in detail in section \ref{sec:cb}.
	
	\textit{Case F: Dynamic amplitude and phase modulation, and OBI.} In this test, the fundamental signal with dynamic amplitude and phase modulation is disturbed by an OBI, i.e.,
	\begin{equation}
		\begin{split}
			x(t)&=(1+0.1\cos(2\pi f_\text{m}t))\cos (2\pi ft\\&+0.1\cos(2\pi f_\text{m}t-\pi))
			+0.05\cos (2\pi f_\text{i}t+\phi_\text{i}),
		\end{split}
	\end{equation}
	where: $f_\text{m}$ is the modulating frequency for both amplitude and phase, and is set from 0.1 to 2$\,$Hz with a step of 0.1$\,$Hz; the fundamental frequency $f=50\,$Hz.
	
\begin{figure}
		\centering
		\includegraphics[width=0.5\textwidth]{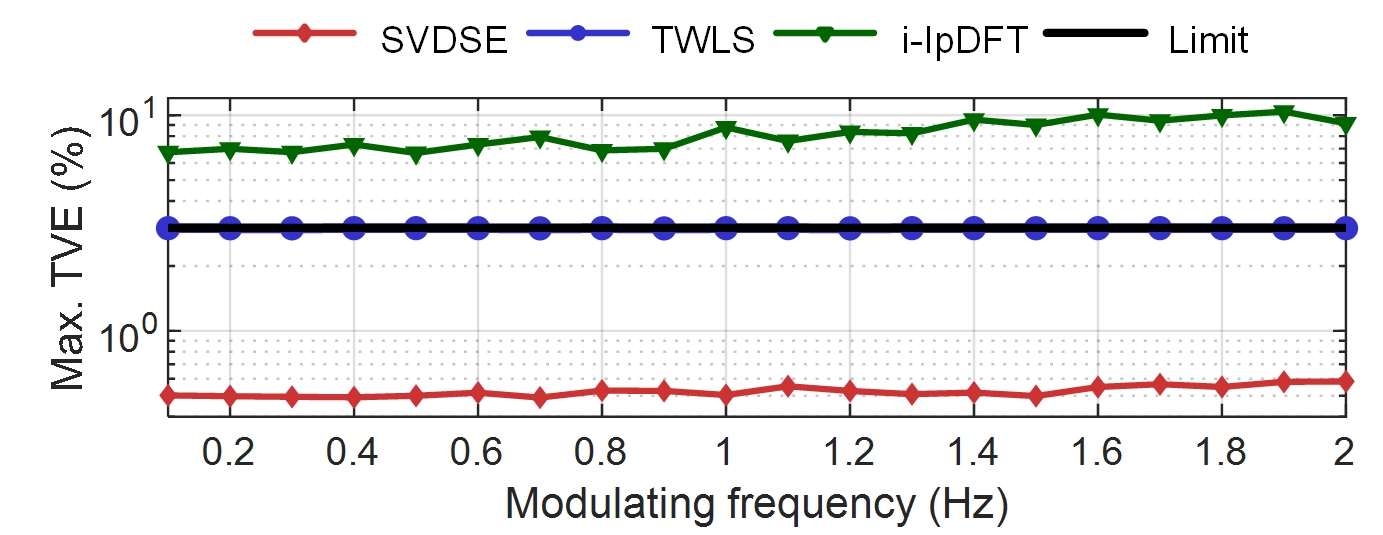}
		\caption{Maximum TVE versus modulating frequency obtained by SVDSE, TWLS, and i-IpDFT algorithms under modulation and OBI condition.}
		\label{fig:07afmod}
	\end{figure}

	As shown in Fig. \ref{fig:07afmod}, for the proposed SVDSE algorithm, the maximum TVE over modulating frequency $f_\text{m}$ up to 2 Hz is always well below the P-class PMU limit (3\%). Similarly, the SVDSE achieves the best performance among the three algorithms.
	
\begin{figure}
		\centering
		\includegraphics[width=0.5\textwidth]{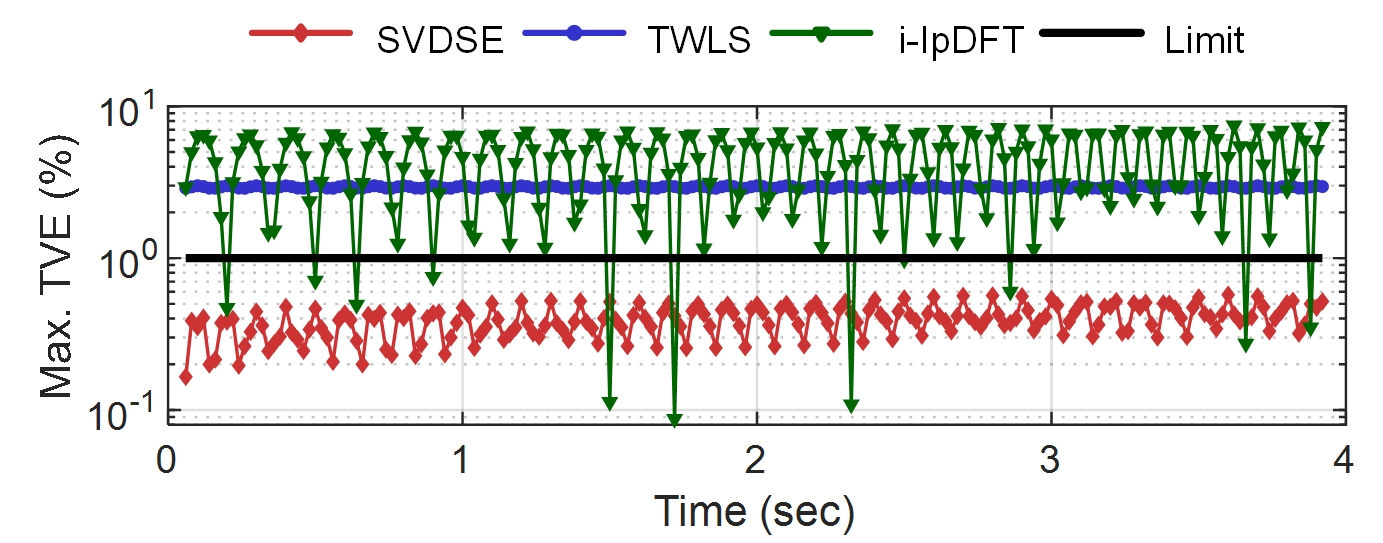}
		\caption{Maximum TVE versus time obtained by SVDSE, TWLS, and i-IpDFT algorithms under frequency ramp and OBI condition.}
		\label{fig:08ramp}
	\end{figure}

	\textit{Case G: Dynamic frequency ramp and OBI.} The ramp fundamental signal with an interharmonic tone is defined as
	\begin{equation}
		x(t)=\cos (2\pi ft+\pi R_\text{f}t^2)+0.05\cos (2\pi f_\text{i}t+\phi_\text{i}),
	\end{equation}
	where the fundamental frequency changes linearly over time with the ramping slope $R_\text{f}=1$ Hz/s (the test result is similar when $R_\text{f}=-1$ Hz/s). Then the fundamental frequency deviation between the real frequency $(f+R_\text{f}t)$ and nominal value increases from -2 to 2 Hz in 4 seconds. Again, the maximum TVE obtained by the proposed SVDSE shown in Fig. \ref{fig:08ramp} meets the P-class limit requirement (1\%) and is the lowest in most shown conditions among the three algorithms.
	
	\textit{Case H: Step change.} In the step change test, the filter response to a signal step change of amplitude or phase is measured. The test signal is given as
	\begin{equation}
		x(t)=(1+k_{\text{a}}\varepsilon(t))\cos(2\pi ft+k_{\text{f}}\varepsilon(t)),
	\end{equation}
	where $\varepsilon(t)$ is the step function, i.e., $\varepsilon(t)=0$ when $t<0$ and $\varepsilon(t)=1$ with $t\ge0$; $k_{\text{a}}=\pm 0.1$ and $k_{\text{f}}=\pm\pi/18$ are magnitude change and phase change recommended by the Standard. We note the sign of $k_{\text{a}}$ and $k_{\text{f}}$ presents only the step direction (rising or falling), and hence in the test, $k_{\text{a}}=0.1$ and $k_{\text{f}}=-\pi/18$ are adopted. The test is done with independent configurations of $k_{\text{a}}$ and $k_{\text{f}}$, yielding two test signals, i.e., 1) $k_{\text{a}}=0.1$, $k_{\text{f}}=0$, and 2) $k_{\text{a}}=0$, $k_{\text{f}}=-\pi/18$. The fundamental frequency $f= 50\,$Hz.
	
\begin{figure}
		\centering
		\includegraphics[width=0.5\textwidth]{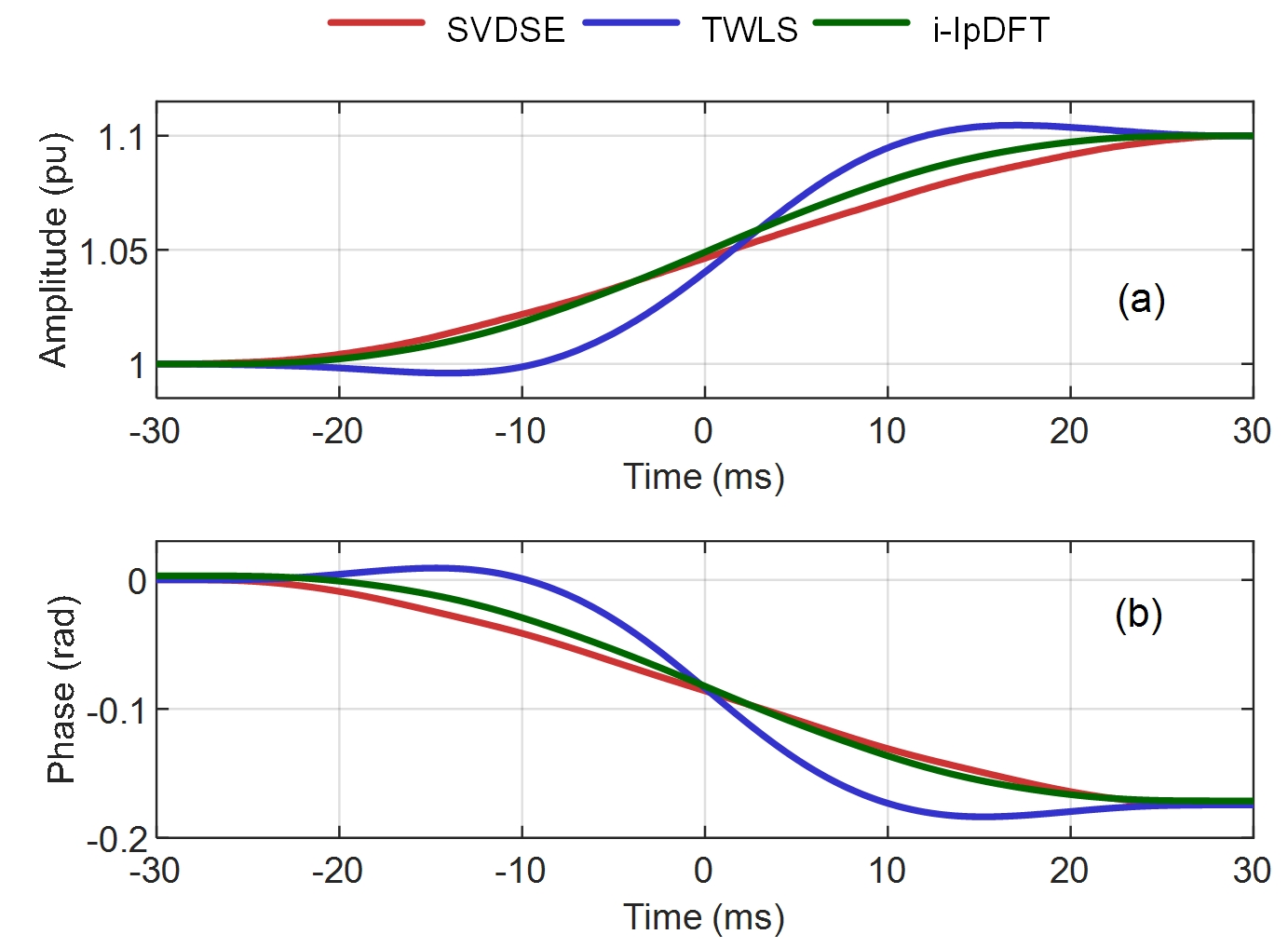}
		\caption{The (a) amplitude and (b) phase responses versus time obtained by SVDSE, TWLS, and i-IpDFT algorithms under amplitude and phase step tests. }
		\label{fig:09setp}
	\end{figure}

	The amplitude and phase step responses obtained by the three algorithms are shown in Fig. \ref{fig:09setp}. It can be seen that the results obtained by the SVDSE and i-IpDFT contain no overshoot or undershoot. Besides, according to the TVE limit set to 1\% in the Standard, all the three algorithms' response times shown in table \ref{tab:02} can meet the P-class requirement, i.e., under two nominal cycles \cite{relays2018118}.
	
	\begin{table}[tp!]
		\centering
		\caption{The response time of TVE obtained by SVDSE, TWLS and i-IpDFT algorithms. The results are expressed as nominal cycles.}
		\setlength{\tabcolsep}{1.8mm}
		\renewcommand\arraystretch{1.3}
		\begin{tabular}{ccccccc}
			\hline
			step set & SVDSE & TWLS & i-IpDFT & Limit\\
			\hline
			amplitude 	&1.73	&0.71	&1.81   &2  \\
			\hline
			phase 	&1.99	&0.86	&1.74   &2  \\
			\hline
		\end{tabular}
		\label{tab:02}
	\end{table}
	
	\subsection*{2) Experimental tests}
	\textit{Case I: OBI of different frequency and natural noise.}
	
	To verify the theoretical analysis and test the robustness of the proposed SVDSE to interharmonic tones and natural noise, case B, where the proposed SVDSE performs the worst in the numerical tests, is carried out experimentally. The test signal has a form of (\ref{eq:12}), containing the fundamental and only one interharmonic tone with a different fixed frequency for each test. The test signal is generated by a programmable signal generator (Tektronix AFG 31000). The signal generator can produce arbitrary waveforms corresponding to the input signal functions. In this experiment, the fundamental amplitude is chosen 1\,V, and the fundamental frequency $f=50\,$Hz; the interharmonic amplitude is 10\% of the fundamental and its frequency ramps within two frequency intervals [10, 25]$\,$Hz and [75, 100]$\,$Hz with a step of $5\,$Hz; the phases of the fundamental $\phi_1$ and the interharmonic $\phi_\text{i}$ are set randomly within $[-\pi, \pi]$. The signal is sampled by a digital voltmeter (Keysight 3458A) and the sampling frequency is set to $5000\,$Hz. The samples are processed by the SVDSE, TWLS, and i-IpDFT algorithms to obtain the dynamic synchrophasor and its error. In the analysis, the reporting rate, time window length, Taylor expansion order, values of the multipliers, iterative numbers, and judgment threshold are the same as the numerical settings in case B. 
	
\begin{figure}
		\centering
		\includegraphics[width=0.5\textwidth]{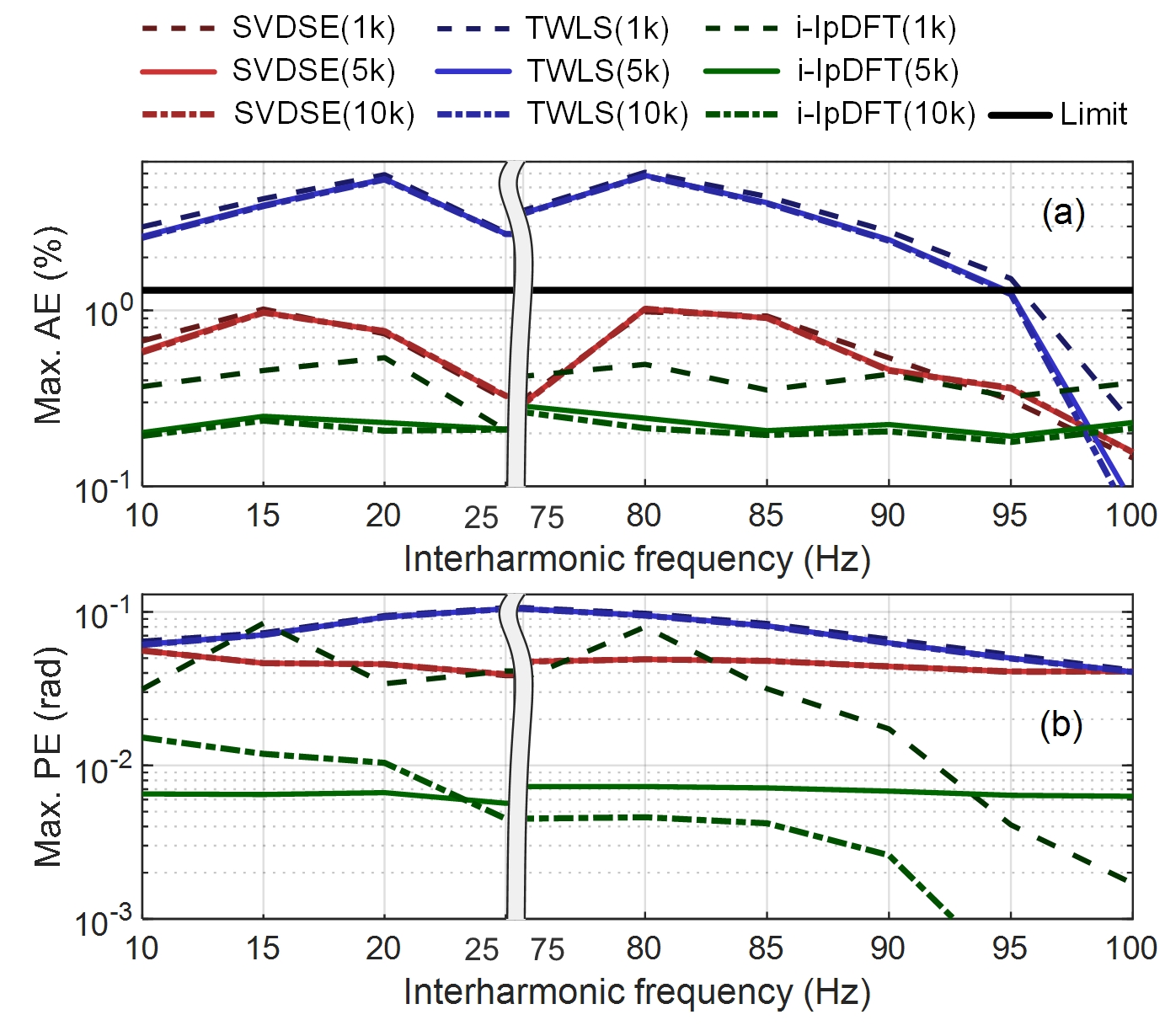}
		\caption{The maximum (a) AE, (b) PE versus interharmonic frequency obtained by SVDSE, TWLS, and i-IpDFT algorithms under the experimental test. }
		\label{fig:11exp}
	\end{figure}

	The measurement errors analyzed by three algorithms are shown in Fig. \ref{fig:11exp}. Subplot (a) shows the maximum amplitude error (AE), presenting the relative amplitude difference between the measurement and the set value. Since the real phase of the signal sample is unknown, it is impossible to obtain the absolute phase error. We use the lowest theoretical error in case B, i.e., the i-IpDFT algorithm, as the reference, and show the relative phase error (PE) change between different algorithms, which is shown in subplot (b). The PE of i-IpDFT in this subplot is set as the simulation error dealing with the signals of the same settings. Although in this case, the TVE is no longer applicable, both the AE curve and the PE curve (relative) can well follow the trend of the theoretical analysis in Fig. \ref{fig:03interf}. Especially, the AE is well below the limit (1\%), and hence verify the effectiveness of the proposed SVDSE algorithm.
	
	Another test on the influence of sampling frequency is also carried out in the above experiment. The same setup with two additional sampling frequencies, respectively 1kHz and 10kHz, is applied, and the PE is obtained with the same phase reference as last test, i.e., $f_\text{s}=5$\,kHz. The performance of the three algorithms is attached in Fig. \ref{fig:11exp}. For both AE and PE of the proposed SVDSE and the TWLS algorithm, a minor difference is obtained for the three sampling frequencies, $f_\text{s}=1$\,kHz, $f_\text{s}=5$\,kHz, and $f_\text{s}=10$\,kHz. Whereas for the i-IpDFT algorithm, the AE and PE increase and get scattered at $f_\mathrm{s}=1$\,kHz.
	
	\section{Discussions of some issues of SVDSE}
	\label{sec5}
	
	Both the theoretical analysis and the above tests show that by changing the singular values in the proposed SVDSE algorithm, the constructed multipliers improve significantly the SE ability to suppress interharmonic interference. As a result, the SVDSE performs better than TWLS and i-IpDFT in most test examples demonstrated. For the TWLS, the window function differing from the rectangular enlarges the transition band gain, and hence causes a weak suppression of interharmonic interference. For the i-IpDFT, the threshold value is required to distinguish between the dynamic behavior (effective signal) and an interharmonic tone to be eliminated \cite{dervivskadic2017iterative}. As a result, the algorithm is not suitable for dealing with interharmonic interfering of small amplitude, e.g., below 8\% of the fundamental tone shown in Fig. \ref{fig:02intera}. By contrast, the main limitation of the proposed SVDSE is that the TVE increases with the interharmonic amplitude. It works well with interharmonic tones below 42\% of the fundamental. When the interharmonic is over 42\%, the TVE of the proposed method may exceed the out-of-band test limit (1.3\%) in the Standard.
	
	For a good SE design, the reference frequency adjustment can not diverge, and the computational speed should be fast, which ideally should meet the requirement for real-time applications. These two aspects are discussed in the following. 
	
	\subsection*{1) Discussion of frequency accuracy of SVDSE}
	The maximum FEs shown in table \ref{tab:03} obtained by the proposed SVDSE in the above typical tests verify that the used frequency adaption method is effective and the obtained frequency does not diverge. The reasons are: 1) Among these tests, the maximum FE is 0.35 Hz appearing in case B where the fundamental signal with a frequency deviation of -2 Hz is disturbed by an interharmonic tone. As seen in Fig. \ref{fig:01Filter}, the gain at the frequency deviation 0.35 Hz is only -0.00164 dB, which is small enough to keep high accuracy when using SVDSE to obtain fundamental phasor; 2) The maximum TVEs of SVDSE are always under the limit among these tests, and the case remains true especially in some dynamic conditions, e.g., case F where signal amplitude and phase are modulated simultaneously and case G where signal frequency ramps over time at slope 1 Hz/s within a wide range [-2, 2]\,Hz.
	
	\begin{table}[tp!]
		\centering
		\caption{The maximum error of frequency obtained by SVDSE algorithm.}
		\setlength{\tabcolsep}{2mm}
		\renewcommand\arraystretch{1.3}
		\begin{tabular}{cccccccc}
			\hline
			signal setting & Case A & Case B & Case C & Case D \\
			\hline
			FE (Hz) 	&0.269 &0.350 &0.172 &0.184  \\
			\hline
			signal setting & Case E & Case F & Case G & Case I\\
			\hline
			FE (Hz) 	 &0.182 &0.186 &0.184 &0.239 \\
			\hline
		\end{tabular}
		\label{tab:03}
	\end{table}
	\subsection*{2) Computational burden analysis}
	\label{sec:cb}
	The main real-time computation of the proposed SVDSE algorithm (shown in table \ref{tab:01}) lies in the solution of filter FIR coefficients in (\ref{eq:10}) and synchrophasor derivatives in (\ref{eq:3}). Supposing that the matrices $\textbf{V}\Sigma^{-1}$, $\textbf{U}$, and the multipliers $m_{1k}(1\le k \le K+1)$ are stored in the memory in advance, then the two real-time computation parts will perform $2N(N-1)+2(K+1)(5K+1)$ real additions, $2N^2+2(K+1)(2N+3K+3)$ real multiplications, and a matrix inversion with time complexity $O((K+1)^3)$. As used in the tests, i.e., $K=2, N=299$, the real-time computational burden includes 178270 real additions and 182444 real multiplications, plus an ignorable time complexity of matrix inversion because $K$ is small. The above tests show that both SVDSE and i-IpDFT can effectively suppress OBI under limited window length. To compare the computation speed, an OBI test where the signal contains the fundamental and one fixed interharmonic tone is applied to both algorithms. The other settings are the same as case A. The test is implemented for 10000 runs by Matlab R2020a on a computer with a 16 GB RAM and a 3.2 GHz processor. The results show the average execution times for one frame of the proposed SVDSE and i-IpDFT are about 5.9 ms and 175.1 ms, respectively. It is shown that the response latency of the SVDSE is much lower than that of i-IpDFT and the most crucial Standard limit, i.e., 10 ms, when reporting rate is 100 frames/sec \cite{relays2018118}. Besides, it is ensured that the computational burden of the proposal does not increase with the number of interference components or iterative numbers as the i-IpDFT.
	
	\section{Conclusion}
	\label{sec6}
	To avoid the misoperations of various protection schemes, fault diagnosis, and control applications caused by interharmonic interference, the synchrophasor estimation methods for P-class PMU with suppression of interharmonic tones have attracted increasing attention. This paper introduces singular value decomposition into the least square algorithm based on Taylor functions and further proposes an optimal solution to design a dynamic synchrophasor filter with good attenuation in the transition band. The proposed algorithm is tested under various signals, such as the fundamental with frequency deviation, dynamic modulations contaminated by OBI, noise, and harmonic. Both the numerical and experimental test results show that the proposed SVDSE algorithm can meet the Standard's requirements under these test conditions. Moreover, the computational burden analysis illustrates that the proposed SVDSE algorithm satisfies the real-time reporting requirement even at the largest rate. The idea that applying singular value decomposition to decompose the matrix formed by filter FIR coefficients and hence improving the filter transition band performance is firstly proposed in this paper. A similar process can be extended to design FIR filters with enhanced attenuation to the interharmonic interference for dynamic synchrophasor derivatives.
	
	\section*{appendix}
	In this appendix, we analyze the characteristics of the first row elements in right singular matrix $\textbf{V}$, i.e., $v_{1(2a+1)}\neq 0\enskip(0\le a\le\lfloor K/2\rfloor)$ and $v_{1(2a)}=0\enskip(1\le a\le\lfloor(K+1)/2\rfloor)$ where $\lfloor\cdot\rfloor $ represents round down operation; $K$ is the Taylor expansion order of the signal model.
	
	According to the SVD of $\textbf{B}_K$ in (\ref{eq:5}), the eigenvalue decomposition of $\textbf{B}_K^\text{T}\textbf{B}_K$ can be written as \cite{kalman2018svd} 
	\begin{align}
		\textbf{B}_K^\text{T}\textbf{B}_K=\textbf{V}\Sigma^{\text{T}}\textbf{U}^\text{T}\textbf{U}\Sigma\textbf{V}^\text{T}=\textbf{V}\Sigma^{\text{T}}\Sigma\textbf{V}^\text{T}.
		\label{eq:A1}
	\end{align}
	Eq. (\ref{eq:A1}) shows that the right singular matrix $\textbf{V}$ of $\textbf{B}_K$ is the eigenvector matrix of $\textbf{B}_K^\text{T}\textbf{B}_K$. Then, according to the relationship between eigenvectors and eigenvalues of a Hermitian matrix found by \cite{denton2021eigenvectors}, the first row elements in the matrix $\textbf{V}$, i.e., eigenvectors of the Hermitian matrix $\textbf{B}_K^\text{T}\textbf{B}_K$, can be obtained by 
	\begin{align}
		v_{1i}=\frac{\prod\limits_{j=1}^{K}\{\lambda_i(\textbf{B}_K^\text{T}\textbf{B}_K)-\lambda_j(\textbf{M}_1)\}}{\prod\limits_{j=1,j\neq i}^{K+1}\{\lambda_i(\textbf{B}_K^\text{T}\textbf{B}_K)-\lambda_j(\textbf{B}_K^\text{T}\textbf{B}_K)\}},1\le i \le (K+1),
		\label{eq:A2}
	\end{align}
	where the minor $\textbf{M}_1$ is formed by removing the first row and column of $\textbf{B}_K^\text{T}\textbf{B}_K$ shown in (\ref{eq:A3}); $\lambda$ represents the eigenvalue of $\textbf{B}_K^\text{T}\textbf{B}_K$ and $\textbf{M}_1$. Eq. (\ref{eq:A2}) shows the first row elements in $\textbf{V}$ can be obtained by the eigenvalues of $\textbf{B}_K^\text{T}\textbf{B}_K$ and $\textbf{M}_1$, whose characteristics will be analyzed in the following text.
	
	As seen from (\ref{eq:A3}), when $(i+j)$ is even, $(\textbf{B}_K^\text{T}\textbf{B}_K)_{ij}=2\sum\limits_{n=1}^{N_h}\frac{(nT_\text{s})^{(i+j-2)}}{(i-1)!\cdot (j-1)!}\quad$, otherwise $(\textbf{B}_K^\text{T}\textbf{B}_K)_{ij}=0\enskip(1\le i,j \le K+1)$ except that $i=j=1,\enskip(\textbf{B}_K^\text{T}\textbf{B}_K)_{11}=2N_h+1$. Similarly, when $(i+j)$ is even, $(\textbf{M}_1)_{ij}=2\sum\limits_{n=1}^{N_h}\frac{(nT_\text{s})^{(i+j)}}{i!\cdot j!} \quad$, otherwise $(\textbf{M}_1)_{ij}=0\enskip(1\le i,j \le K)$. Therefore, matrices $\textbf{B}_K^\text{T}\textbf{B}_K$ and $\textbf{M}_1$ have similarity in the composition and have full rank of $(K+1)$ and $K$, respectively. When using Gauss elimination to obtain the eigenvalues of matrices  $\textbf{B}_K^\text{T}\textbf{B}_K$ and $\textbf{M}_1$, it can be obtained that $\lambda_{(2a)}(\textbf{B}_K^\text{T}\textbf{B}_K)=\lambda_{(2a-1)}(\textbf{M}_1)\enskip(1\le a\le\lfloor(K+1)/2\rfloor)$. As a result, when $i=2a$, there will be a numerator term in (\ref{eq:A2}) that equals to zero and can not be reduced by any denominator term, i.e., $(\lambda_{(2a)}(\textbf{B}_K^\text{T}\textbf{B}_K)-\lambda_{(2a-1)}(\textbf{M}_1))$, resulting $v_{1(2a)}=0\enskip(1\le a\le\lfloor(K+1)/2\rfloor)$. Otherwise, when $i=2a+1$, there is no such term in numerator and $v_{1(2a+1)}\neq0\enskip(0\le a\le\lfloor K/2\rfloor)$.
	
	\begin{align}
		\textbf{B}_K^\text{T}\textbf{B}_K=
		\begin{pmatrix}
			N & 0 & 2\sum\limits_{n=1}^{N_h}\frac{(nT_\text{s})^2}{2!}  &  \ldots \\
			0 & 2\sum\limits_{n=1}^{N_h}(nT_\text{s})^2 & 0 &  \ldots\\
			2\sum\limits_{n=1}^{N_h}\frac{(nT_\text{s})^2}{2!} & 0 &2\sum\limits_{n=1}^{N_h}\frac{(nT_\text{s})^4}{2!\cdot2!} &  \ldots\\
			\vdots & \vdots & \vdots & \ddots 
		\end{pmatrix}.
		\label{eq:A3}
	\end{align}

\bibliographystyle{IEEEtran}

\end{document}